\RequirePackage{fix-cm}
\documentclass[smallcondensed]{svjour3}     % onecolumn (ditto)
\smartqed  % flush right qed marks, e.g. at end of proof
\usepackage{graphicx}
\usepackage{amsmath}
\usepackage{epsfig}
\usepackage{stmaryrd}
\usepackage{amsfonts}
\usepackage{amssymb}
\usepackage{wasysym}

\begin{document}

\title{An analysis of a regular black hole interior model%\thanks{Grants or other notes
}

\author{Daniela P\'erez \and Gustavo E Romero \and Santiago E Perez-Bergliaffa}

%\author{First Author         \and
%        Second Author %etc.
%}


%\institute{F. Author \at
%              first address \\
%              Tel.: +123-45-678910\\
%              Fax: +123-45-678910\\
%              \email{fauthor@example.com}           %  \\

\institute{Daniela P\'erez \and Gustavo E Romero \at
              Instituto Argentino de Radioastronom{\'\i}a, Camino Gral Belgrano Km 40\\
              C.C.5, (1984) Villa Elisa, Bs. As., Argentina\\ 
              Tel.: +54-221-482-4903\\
              Fax: +54-221-425-4909\\
              \email{danielaperez@iar.unlp.edu.ar}           %  \\
           \and
           Gustavo E Romero \at
           Facultad de Ciencias Astron\'omicas y Geof{\'\i}sicas, Universidad Nacional de La Plata,\\ Paseo del Bosque s/n, 
					 CP (1900), La Plata, Bs. As., Argentina\\
           \email{romero@iar.unlp.edu.ar}
           \and
           Santiago E Perez-Bergliaffa \at
           Departamento de F\'{\i}sica Te\'orica, Instituto de F\'{\i}sica, Universidade do Estado do Rio de Janeiro,
           Rua S\~ao Francisco Xavier 524, Maracan\~a Rio de Janeiro - RJ, Brasil, CEP: 20550-900}

\date{Received: date / Accepted: date}

\maketitle

\begin{abstract}
We analyze the thermodynamical properties of the regular static and spherically symmetric black hole interior model presented by Mboyne and Kazanas. 
Equations for the thermodynamical quantities valid for an arbitrary density profile are deduced, and from them we show that the model is thermodynamically unstable. Evidence is also presented pointing to its dynamical instability. The gravitational entropy of this solution based on the Weyl curvature conjecture is calculated, following the recipe given by Rudjord, Gr$\varnothing$n and Sigbj$\varnothing$rn, and it is shown to have the expected behavior. 

\end{abstract}

\section{Introduction}

The existence of singularities in some solutions of General Relativity 
is one of the most important unresolved issues in the classical description of
the gravitational field. Singularities are an undesirable feature 
of any theory of gravitation:
they can
be naturally considered as a source of lawlessness (see for instance \cite{earman}),
because the space-time description
breaks down ``there'', and physical laws presuppose space-time. 
There are several ways in which singularities can be avoided 
(see \cite{physrep} for the case of Cosmology), both at the classical and at the quantum level. Since there is no widely-accepted theory of quantum gravity, the issue of singularities in General Relativity has been addressed from classical considerations
in various fashions. One particular approach is the introduction of a limiting curvature hypothesis, based on that, at a singularity, invariants constructed with the Riemann tensor generally diverge. 
Hence, a suitable way to eliminate singularities would be to 
impose a limiting principle on the 
curvature, associated with a fundamental scale (it may be the Planck scale).  
This hypothesis 
(first proposed in \cite{markov})
is implemented by demanding that 
any solution of the field equations must
reduce to a definite nonsingular solution (for instance to a de Sitter space-time)
when some of the invariants attain their limiting values, 
in such a way that the strong energy condition (one of the hypothesis of the singularity theorems \cite{s1}\cite{s2}\cite{s3}) is violated.

A theory based on this hypothesis has been implemented by modifying the dynamics of General 
Relativity with the addition of nonlinear terms in Ref. \cite{bran}. However, 
there is another approach, suggested by Gliner \cite{gli}, that consists of assuming that the micro-physics of high-density
matter is such that a phase transition must occur leading any system under such extreme conditions to the de Sitter geometry. 
Models of this type were also considered by Sakharov \cite{sak}, Zeldovich
\cite{zel}, and Bardeen \cite{baar}. In these models the space-time metric behaves as the Schwarzschild metric for large radii and, as de Sitter's towards the core of the object. Regular black hole models with direct matching of the de Sitter interior and the Schwarzschild exterior have been proposed by Poisson and Israel \cite{poi}, Frolov et al. \cite{frolov89,frolov90}, and Dymnikova \cite{dy2003}.

An exact solution of the Einstein field equations containing at the core a de Sitter fluid was found by Dymnikova \cite{dy}. 
This solution represents a vacuum nonsingular black hole. The stress-energy tensor which is the source of curvature ``describes a smooth transition from the standard vacuum state at infinity to (an) isotropic vacuum state at $r\rightarrow 0$ through (an) anisotropic vacuum state in (the) intermediate region [...]. At the present such a transitional material cannot be described starting from any fundamental theory describing reality at the microscopic level". 
 
This anisotropic vacuum state was replaced by 
Mbonye and Kazanas (MK) \cite{mk} with a region filled with 
matter under both radial and tangential pressures, described
by an equation of state (EOS) that relates the radial pressure to the density
and smoothly reproduces the de Sitter's spacetime behavior near $r = 0$, 
tending to a polytrope at larger $r$, low $\rho$ values\protect\footnote{Different types of regular black hole solutions were reviewed recently in \cite{ze}, where 
regular charged black holes, with the exterior described by the Reissner-Nordstrom solution were discussed as well. Regular black holes with no center at all have been discussed by Bronnikov et al. \cite{bro2006,bro2007}.}.

The main goal of the present work is to study the MK model  for a regular black hole in which the singularity is avoided precisely by the second method mentioned above. In particular, 
a detailed study of the thermodynamical aspects of the matter that is the source
of the curvature and of the gravitational field will be given\protect\footnote{Other features of the model were studied in \cite{mk2}. Thermodynamics of exotic matter is discussed, among others, by: \cite{ex1}\cite{ex2}\cite{ex3}.}. It will be shown that the solution is thermodynamically unstable, and we shall present evidence supporting the dynamical instability of the model. The issue of the
gravitational entropy of this model will be also addressed, following the 
ideas presented in \cite{Gron} (see also Romero et al. \cite{us}).

We shall begin in Sect.\ref{model} with a short review of the regular black hole solution advanced in \cite{mk}. Its thermodynamical properties will be discussed in Sect.\ref{terprop}, and the stability will be analyzed in Sect.\ref{stab}. In Sect.\ref{gravent}, the gravitational entropy of the MK solution will be calculated. These results will be analyzed in Sect.\ref{disc}.

\section{\label{model} A model of a regular black hole interior}

\begin{figure}[b]
\begin{center}
\includegraphics{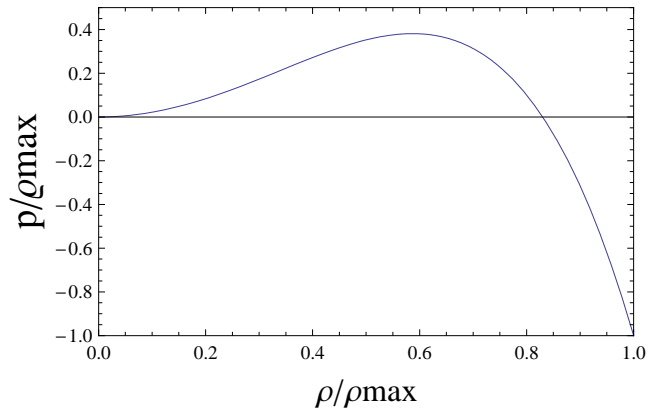}% Here is how to import EPS art
\end{center}
\caption{Plot of (\ref{es}), that gives the relation between the radial pressure and the density.}
\label{es}
\end{figure}

The model introduced in \cite{mk} represents a regular static black hole, 
with a matter source that smoothly goes from a de Sitter behavior near the origin to Schwarzschild's space-time outside the object. 
A space-time metric well-adapted to examine the properties of this system is
\cite{mk2}:
\begin{equation}
ds^{2} =-B(r)dt^{2}+\left(1-\frac{2m(r)}{r}\right)^{-1}dr^{2}+r^{2}(d\theta^{2}+\sin^{2}\theta d\phi^{2}).
\label{met}
\end{equation}
The EOS proposed by Mbonye and Kazanas \cite{mk} is given by:
\begin{equation}\label{3}
p_{r}(\rho ) = \left[\alpha-(\alpha+1)\left(\frac{\rho}{\rho_{\textsf{max}}}
\right)^{2}\right]\left(\frac{\rho}{\rho_{\textsf{max}}}\right)\rho,
\end{equation}
where $p_r$ is the radial pressure (to be distinguished from the tangential pressure, which is needed in these models to satisfy the Tolman-Oppenheimer-Volkoff (TOV) equation\footnote{As shown in \cite{visser}, in order to satisfy the  TOV equation, the anisotropic pressure must be introduced when dealing with static spherically symmetric space-times and perfect fluids with negative pressures in some interval of the radial coordinate as a source.}), and $\alpha=2.2135$ \protect\footnote{This value of $\alpha$ is chosen to yield a sound speed bounded by the speed of light \cite{mk}.}. 
Such an equation describes the behavior of matter that changes smoothly from normal to a core of ``exotic fluid'' with an EOS that approaches $p_r=-\rho$ when $r \rightarrow  0$ (see Figure \ref{es}). It follows that in order to avoid the presence of the singularity at $r=0$, the strong energy condition is violated in the core of the object, whereas the weak energy condition and dominant energy condition are everywhere satisfied \cite{mk}.

MK chose the density profile suggested by Dymnikova to solve Einstein's equations at small $r$\cite{dy}\footnote{MK assumes implicitly that $\rho=0$ for $r>R$, although this is not strictly correct. The cutoff in the density profile, however, is sufficiently strong as to consider this approximation correct from any practical point of view.}:
\begin{equation}\label{7}
\rho(r)=\rho_{\textsf{max}}e^{-8\frac{r^{3}}{R^{3}}},
\end{equation}
where $R^{3}=8R_{\rm S}r_{0}^{2}$, $R_{\rm S}$ is the Schwarzschild radius, 
\begin{equation}
r_{0}=\left(\frac{3}{8\pi \rho_{\textsf{max}}}\right)^{1/2},
\end{equation}
and $\rho_{\rm max}$ is of the order of the Planck density.  
It follows from (\ref{7}) that 
(almost) all the mass is contained inside a sphere of radius $R$ well within the black hole, which corresponds to the surface of the object inside the horizon. The density undergoes a smooth transition from the de Sitter state at the center to the vacuum state for $r> R$, with an intermediate region 
of non-inflationary material, a situation that was anticipated in \cite{poi}.

From Eq. \ref{met}, when $r\rightarrow \infty$, $m(r)\rightarrow M$, and the external metric ($r>>R_{\rm S}$ becomes asymptotically Schwarzschild.

The structure of the black hole interior is characterized by the followings regions, as described in \cite{mk}:
\begin{enumerate}
\item Exotic matter region:
\begin{equation}
0\leq r \leq r_{*}=0.4R,
\end{equation}
where $R$ is the radius of the compact object inside the horizon. At the core of this region the metric is de Sitter.\\
\item Normal matter region:
\begin{equation}
0.4R\leq r \leq R.
\end{equation}
This region is supported against collapse by the de Sitter core.\\
\item Schwarzschild vacuum region:
\begin{equation}
R < r < r_{{\rm h}}=2M,
\end{equation}
where $r_{{\rm h}}=R_{\rm S}=2M$ is the radius of the event horizon. In this region the Ricci tensor is null.
\end{enumerate}

We emphasize that the space-time metric is exactly de Sitter at the geometrical center of the object and, for $r>R$ it quickly tends to Schwarzschild.

\begin{figure}[t]
\begin{center}
{\includegraphics{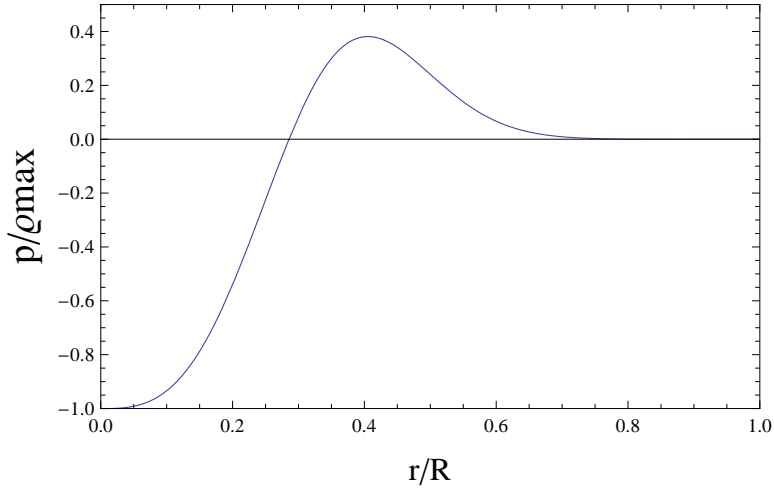}}
\end{center}
\caption{\label{pr}Radial pressure as a function of the radial coordinate inside the regular black hole considered in the text.}
\end{figure}

The plot of the radial pressure as a function of the radial coordinate is shown in Figure \ref{pr}. It can be seen that the pressure follows the equation $p=-\rho$ at the core of the object, and it goes to zero at $r/R=1$, the surface of the matter region. The radial pressure has another zero located at $r/R=0.28$, 
two inflexion points at $r/R= 0.26$ and $r/R= 0.5$, and an absolute maximum at $r/R = 0.4$. The exotic matter (for which the pressure decreases with decreasing $r$) occupies the region $r/R<0.4$ \protect\footnote{Although this profile resembles the qualitative plot for the radial pressure in a gravastar
(see \cite{visser}), it must be noted that the MK model has an horizon, which is absent in gravastars.}.

The solution of the Einstein field equations $G^{\mu}_{\nu}=-8\pi T^{\mu}_{\nu}$, with:
$$T^{\mu}_{\nu}= {\rm diag}(\rho(r), p_r(r), p_\perp (r),p_\perp (r)),$$
for the metric given in (\ref{met}) and the EOS (\ref{3}) takes the form \cite{mk2}:
%\begin{eqnarray}
%B(r)& = &\exp{\left\{\int^{r}_{r_{0}}\frac{2}{r'^{2}}\left[m(r')+4\pi r'^{3}p_{\textsf{r'}}\right] \times{} \nonumber \\
%& &{} \times \left[\frac{1}{\left(1-\frac{2m(r')}{r'}\right)}\right]dr'\right\}},

%\end{eqnarray}
%{\setlength\arraycolsep{2pt}
\begin{eqnarray}
B(r)&=&\exp \int^{r}_{r_{0}}\frac{2}{r'^{2}}\left[m(r')+4\pi r'^{3}p_{r'}\right]\nonumber\\
& &\times\left[\frac{1}{\left(1-\frac{2m(r')}{r'}\right)}\right]dr'.
\label{b}
\end{eqnarray}
The total mass of the object is given by:
\begin{equation}
M = \int^{R}_{0} m(r) dr,
\end{equation}
where:
\begin{equation}
m(r)=4\pi \int^{r}_{0}\rho(r')r'^{2}dr'.
\label{m}
\end{equation}

\begin{figure}[pb]
\begin{center}
{\includegraphics{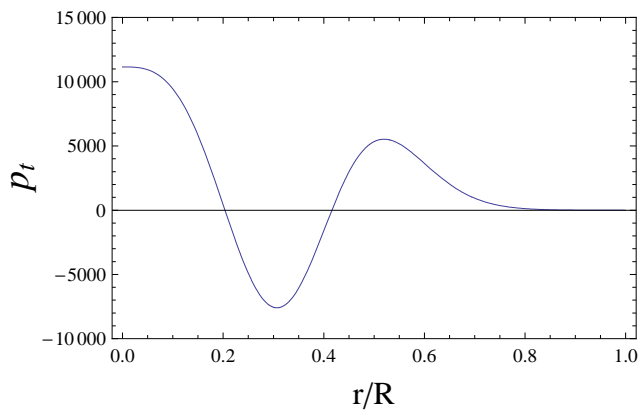}}
\end{center}
\caption{Plot of the tangential pressure as a function of the radial coordinate, given by 
(\ref{pretan}).}
\label{pt}
\end{figure}

Equations (\ref{b}) and (\ref{m}) describe the geometry of the space-time of the  regular black hole introduced in \cite{mk}.
%We can derive from the $g_{\rm rr}$ component of the metric that this model presents an event horizont at $R$=2$M$. Here, $R$ is the radius at which the total mass $M$ of the black hole is contained.
Since the EOS and the density as a function of $r$ were specified in \cite{mk},
the tangential pressure in terms of the given quantities follows from the Einstein field equations. Explicitly:
\begin{equation}
p_{\bot} = p_{ r} + \frac{r}{2} p'_{r} + \frac{1}{2} \left(p_{r}+\rho\right)\left[\frac{m(r) + 4 \pi r^{3} p_{r}}{r-2m(r)}\right].
\label{pretan}
\end{equation}
%where $p_{{\textsf r}}$ is given by Eq. (\ref{3}). The formula of the %tangential pressure is a generalization of the Tolman-Oppenheimer-Volkoff %equation.
The behavior of the tangential pressure with $r$ is given in Figure \ref{pt}.
%\begin{figure}[pb]
%\centerline{\psfig{file=ptangencial.eps,width=4.7cm}}
%\vspace*{8pt}
%\caption{Equation of state of the matter fields. \label{es}}
%\end{figure}
It is important to emphasize at this point that the MK model differs from the usual approach (see for instance \cite{horvat}), in the sense that, whereas in the former one the EOS and the explicit form of $\rho (r)$ were given, in the latter
two relations between the quantities $p_r$, $p_\perp$ and $\rho$ are advanced.
As shown in the Appendix, the choice in the MK model has important consequences in the
analysis of the perturbations of the field equations.

%The space-time metric of de Sitter's cosmological solution is:
%\begin{equation}
%ds^2 = %\left(1-\frac{r^2}{{r_{0}}^2}\right)dt^2-\frac{dr^2}{1-\frac{r^2}{{r_{0}}^2}}-r^2\left(d\theta^2+\sin\theta^2 %d\varphi^2\right),
%\end{equation}
%where $r_{0}^2 = 3/\lambda$, being the cosmological constant $\lambda$ related %to the vacuum energy density $\epsilon$ by:
%\begin{equation}
%\epsilon = \frac{\lambda c^{4}}{8\pi G}.
%\end{equation}
%
%Dymnikova assumed a specific form for the stress-energy tensor:
%\begin{equation}\label{dyt}
%T^{0}_{0} = \epsilon_{0} \exp\left(-\frac{r^{3}}{r_{0}^{2}R_{S}}\right),
%\end{equation}
%where $R_{{\rm S}}$ is the radius of the Schwarzschild horizon and $r_{0}$ is %connected to $\epsilon_{0}$ by the de Sitter relation:
%\begin{equation}
%r_{0}^2 = \frac{3c^{4}}{8\pi G \epsilon_{0}}.
%\end{equation}
%
%The solution of the Einstein field equations for the $T^{0}_{0}$ component of %the stress-energy tensor given by Eq. (\ref{dyt}) is regular everywhere and %coincides with the Schwarzschild solution for large radii, and for %$r\rightarrow 0$, it tends to de Sitter space-time.
%
%Mbonye and Kazanas adopted Dymnikova's density profile but replaced the vacuum %energy density $\epsilon_{0}$ by the maximum energy density for the matter %f%ields $\rho_{\rm max}$.

\section{Thermodynamics of the matter inside the horizon}\label{terprop}

Following \cite{mk}, we shall start by assuming that 
the body has reached a static equilibrium configuration (i.e inside the black hole, the pull of the gravitational field is balanced by the repulsion exerted by the exotic matter). It will also be assumed that matter is in thermodynamic equilibrium. As shown below, the results obtained under these hypotheses 
turn out to be incorrect, and hence they 
imply that the 
system is both dynamically and thermodynamically unstable.

The temperature of the matter as a function of the radius can be estimated from
the laws of standard thermodynamics:
\begin{equation}\label{temp}
TdS=d(\rho V)+p_{\textbf{\textbf{r}}}dV,
\end{equation}
where $dV$ is an element of volume in the proper reference frame of the fluid \cite{gravitation}.
%\begin{equation}\label{temp}
%TdS=d(\rho V)+pdV.
%\end{equation}
From here, using $\partial^2S/\partial V \partial T = \partial^2 S / \partial T \partial V$:
\begin{equation}\label{what}
dp_{ r} = \frac{\left(\rho+p_{r}\right)}{T} dT.
\end{equation}
If we replace (\ref{3}) and (\ref{7}) into (\ref{what}) we get:
\begin{equation}\label{15}
\frac{dT}{T} = \frac{2\left[\alpha-2(\alpha+1)(\rho/\rho_{\textsf{max}})^{2}\right]d\rho}{\rho_{\textsf{max}}+\left[\alpha-(\alpha+1)(\rho/\rho_{\textsf{max}})^{2}\right]\rho}.
\end{equation}
We obtain the formula of the temperature as a function of the radial coordinate by integration of (\ref{15}):
%\begin{equation}\label{9}
%dp = \frac{\rho+p}{T}dT.
%\end{equation}
%Then, replacing  into (\ref{9}), we obtain:
\begin{equation}\label{temp1}
\frac{T}{T_{\textsf{sup}}}=\left[1+\alpha e^{-8r^{3}/R^{3}}-(\alpha+1)e^{-24r^{3}/R^{3}}\right]^{4/3}e^{\Xi(r)},
\end{equation}
where $T_{\rm{sup}}$ stands for the temperature of the the matter field at $r = R$, and:
\begin{equation}
\Xi(r)=\frac{2}{3}\int^{e^{-8r^{3}/R^{3}}}_{0}\frac{\alpha d(\rho/\rho_{\textsf{max}})}{1+\alpha(\rho/\rho_{\textsf{max}})-(\alpha+1)(\rho/\rho_{\textsf{max}})^{3}}.
\end{equation}
%\begin{figure}[b]
%  \centering
%  \hfill\begin{minipage}[b]{.45\textwidth}
%  \centering
%\includegraphics[scale=0.9]{Tyes.eps}
%\caption{Temperature as a function of radial coordinate inside the black hole. $T_{\rm{sup}}$ stands for the temperature of the the matter field at $r = R$.}
%\label{figtemp}%
%\end{figure}
%  \end{minipage}~\hfill%
%  \begin{minipage}[b]{.45\textwidth}
%    \centering
\begin{figure}[t]
\begin{center}
\includegraphics{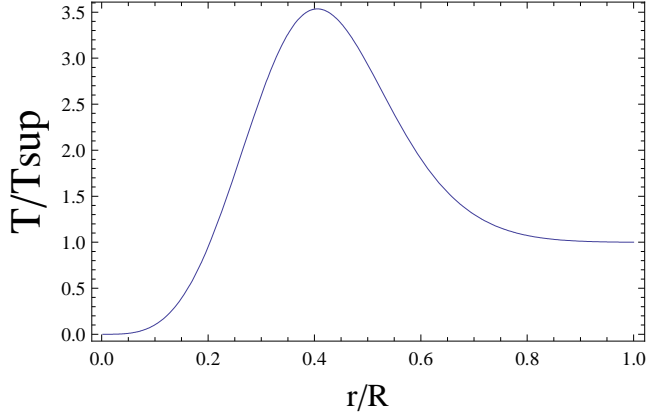}
\end{center}
\caption{Temperature as a function of the radial coordinate inside the black hole, given by (\ref{temp1}). $T_{\rm{sup}}$ stands for the temperature of the the matter field at $r = R$. \label{figtemp}}
\end{figure}
Figure \ref{figtemp} displays the temperature of the matter as a function of the radius. We note that the temperature tends to absolute zero close to the core. As for the pressure, there is only one maximum at $r/R=0.4$. 
The inflexion points of $T(r)$ occur at $r/R= 0.31$ and at $r/R= 0.57$.

An expression for the entropy (up to an additive constant) as a function of the coordinates can be obtained by substituting (\ref{temp1}) in (\ref{temp}): 
\begin{equation}
S\equiv \frac{(\rho+p)}{T}V.
\end{equation} 
\begin{figure}[pb]
\begin{center}
{\includegraphics{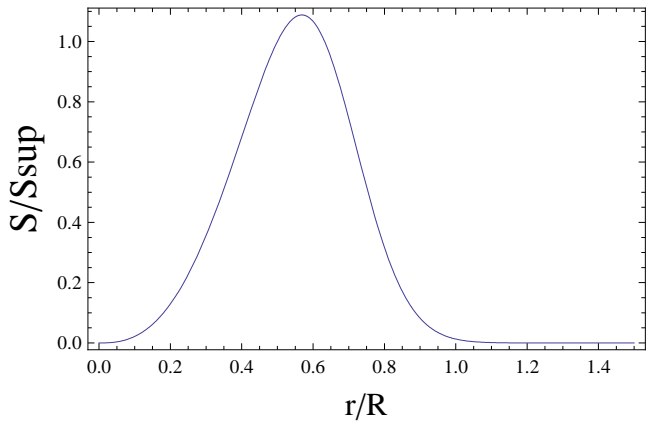}}
\end{center}
\caption{Entropy of the matter field as a function of the radial coordinate inside the black hole. $S_{\rm{sup}}$ stands for the entropy of the the matter field at $r = R$. \label{ENT}}
\end{figure}
The result is plotted in Figure \ref{ENT}. The entropy goes to zero as $r \rightarrow 0$ in accordance to Nerst's theorem, and it has a maximum close to $r/R = 0.4$, in the region of highest density of normal matter.

The entropy density of the matter inside the regular black hole can also be calculated, and is given by:
\begin{equation}\label{dens-termo}
\frac{s}{s_{\textsf{R/2}}}=\frac{(\rho/\rho_{\textsf{max}})[1+\alpha(\rho/\rho_{\textsf{max}})-(\alpha+1)(\rho/\rho_{\textsf{max}})^{3}]^{-1/3}}{0.2076 e^{(2/3)\int^{\rho}_{0}\alpha d\rho[\rho_{\textsf{max}}+\alpha\rho-(\alpha+1)(\rho^{3}/\rho_{\textsf{max}}^{2})]^{-1}}}.
\end{equation}

\begin{figure}[pb]
\begin{center}
{\includegraphics{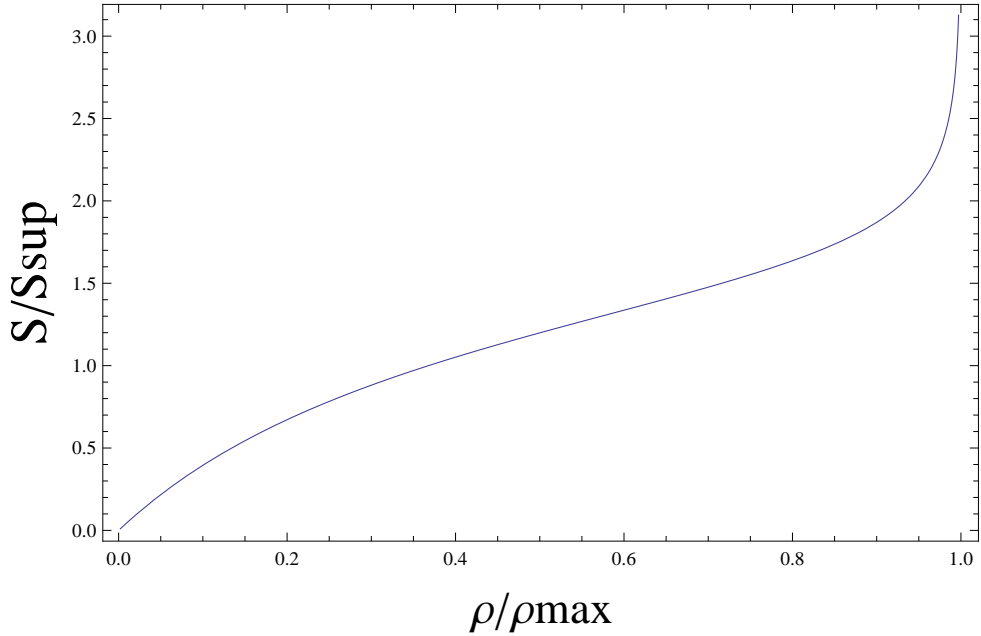}}
\end{center}
\caption{Entropy density of the matter inside the black hole as a function of the density. $S_{\rm{sup}}$ stands for the entropy of the matter field at $r = R$. \label{fig:2}}
\end{figure}

\begin{figure}[t]
\begin{center}
{\includegraphics{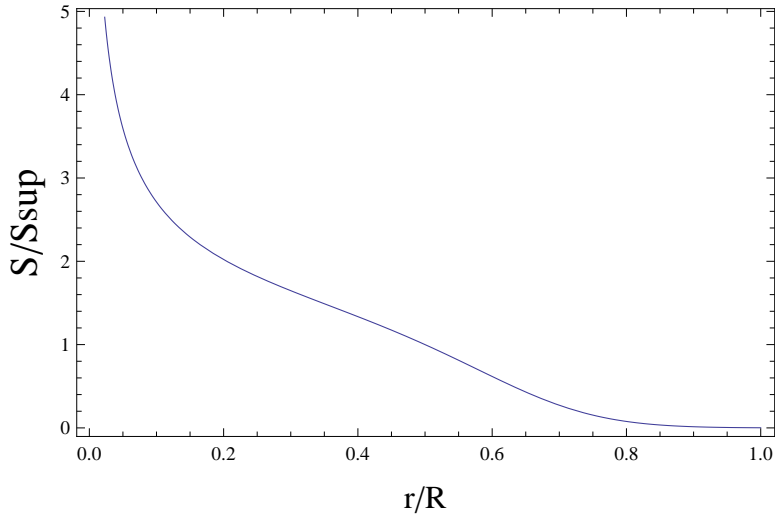}}
\end{center}
\caption{Entropy density of the matter inside the black hole as a function of the radial coordinate. $S_{\rm{sup}}$ stands for the entropy of the matter field at $r = R$. \label{fig:22}}
\end{figure}

The plots of the entropy density as a function of $\rho$ and $r$ are shown in Figures \ref{fig:2} and \ref{fig:22}. It can be seen that the entropy density diverges at the origin as a consequence of the vanishing volume. The entropy density as a function of the density has one inflexion point at $\rho/\rho_{\small{\rm{max}}} = 0.73$ which is located at $r/R= 0.34$.

\begin{figure}[b]
\begin{center}
\includegraphics{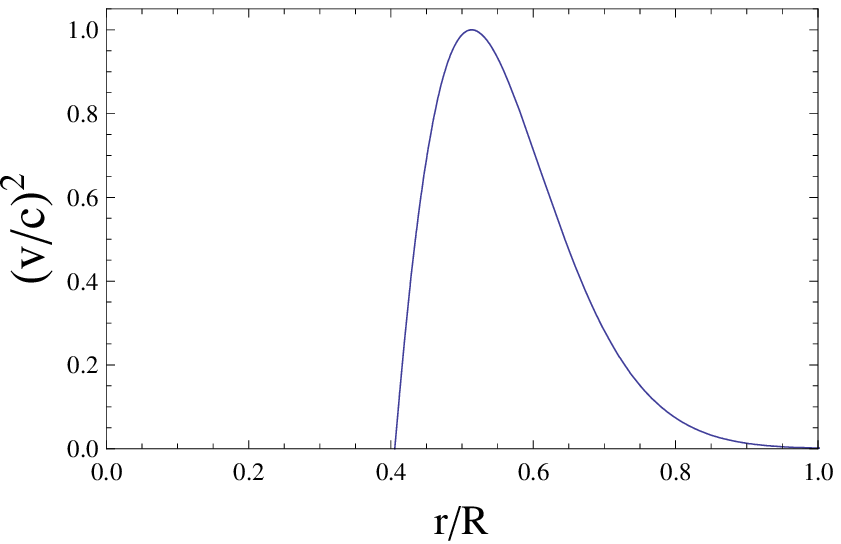}
\end{center}
\caption{Sound speed as a function of the radial coordinate. \label{fig:3}}
\end{figure}

The radial sound speed as a function of the energy density follows from $\ v_r^{2}=dp/d\rho$ (with $c=1$), yielding: 
\begin{equation}\label{1228}
v_r^{2}=2e^{-8r^{3}/R^{3}}\left[\alpha-2(\alpha+1)e^{-16r^{3}/R^{3}}\right].
\end{equation}
The result is shown in Figure \ref{fig:3}. From Figures \ref{pr} and \ref{fig:3}, we see that the sound speed is zero at the value of $r/R$ for 
which the pressure is maximum. In addition, the sound speed takes complex values in the region $r/R<0.4$, in accordance with the negative slope of the equation of state as a function of the density (see (\ref{3}) and Figure \ref{es}). The fact that sound waves do not propagate in the region $r/R<0.4$ is a consequence of the exotic behavior of the fluid there: a variation in pressure causes an expansion rather than a compression. 

\begin{figure}[t]
\begin{center}
{\includegraphics{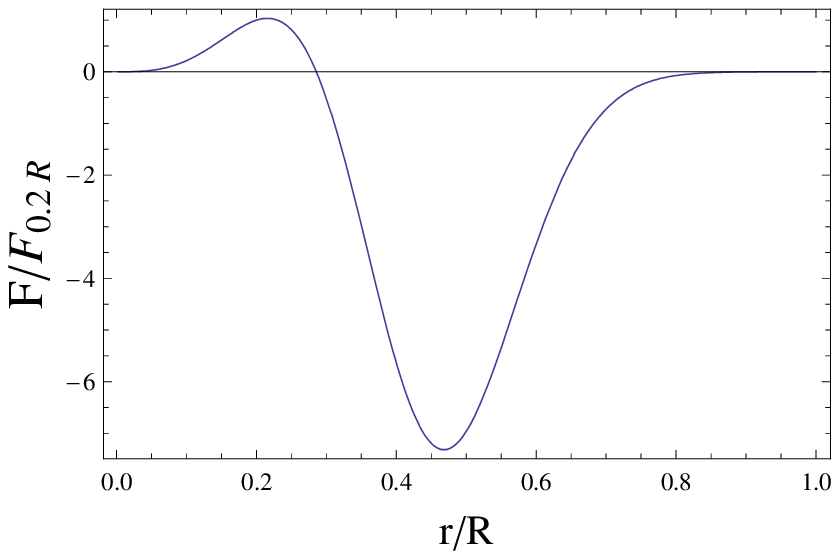}}
\end{center}
%  \end{minipage}~\hfill~\\[0.1pt]
%  \hfill\begin{minipage}[t]{.5\textwidth}
  
%  \end{minipage}~\hfill%
%  \begin{minipage}[t]{.5\textwidth}
\caption{Helmholtz free energy as a function of the radial coordinate. $F_{0.2R}$ stands for the Helmholtz free energy at $r/R =0.2$.}
\label{HEL}%
%  \end{minipage}\hfill~%
\end{figure}

To discuss the thermodynamical equilibrium we shall need below the Helmholtz free energy, given by
$ F=U-TS=-pV$. Explicitly:
\begin{equation}
\frac{F}{F_{0.2R}}=\frac{-1}{0.018}\left[\alpha-(\alpha+1)\left(\frac{\rho}{\rho_{\text{max}}}\right)^{2}\right]\left(\frac{\rho}{\rho_{\text{max}}}\right)^{2}\frac{4\pi}{3}\left(\frac{r}{R}\right)^{3}.
\end{equation}
The plot of the Helmholtz free energy $F$ as a function of the radial coordinate is shown in Figure \ref{HEL}. \\
Using the functions discussed in this section, we shall discuss in the next one
the issue of thermodynamical and dynamical equilibrium.

\section{Equilibrium of the black hole interior configuration}
\label{stab}
\subsection{Thermodynamical equilibrium}

\begin{figure}[t]
\begin{center}
{\includegraphics{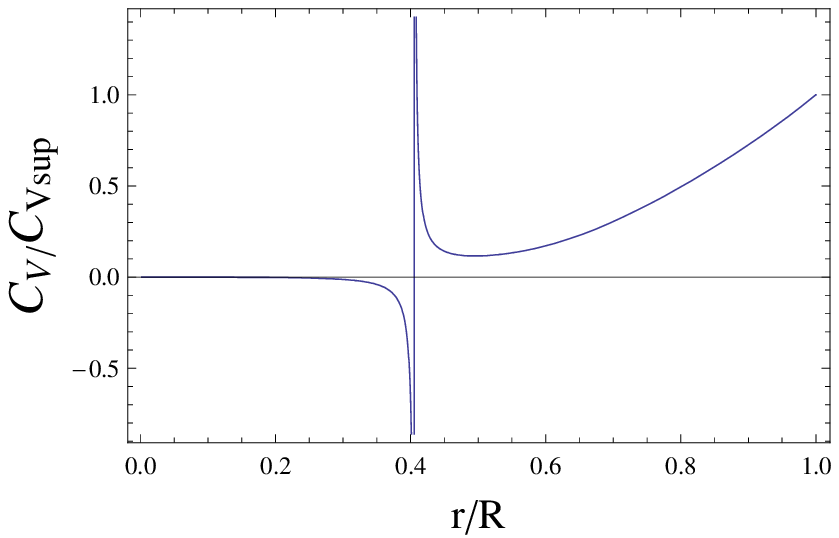}}
\end{center}
%  \end{minipage}~\hfill~\\[0.1pt]
%  \hfill\begin{minipage}[t]{.5\textwidth}
  
%  \end{minipage}~\hfill%
%  \begin{minipage}[t]{.5\textwidth}
\caption{Specific heat at constant volume as a function of the radial coordinate in the black hole interior. $C_{V{\rm sup}}$ stands for the specific heat at constant volume at $r = R$.}
\label{cv}%
%  \end{minipage}\hfill~%
\end{figure}

A condition for a system to be in stable thermodynamical equilibrium is that, for a given value of entropy and volume, the energy must be minimum.
This is equivalent to impose on the 
the specific heat at constant volume the condition:
\begin{equation}
C_{V}>0.\label{ts2}
\end{equation}
%\begin{eqnarray}
%\frac{\partial^{2}U}{\partial S^{2}}\arrowvert_{V} & >0 &,\\
%\frac{\partial^{2}U}{\partial S^{2}}\arrowvert_{V} & = & \frac{\partial %T}{\partial S}\arrowvert_{V}>0.\label{ts1}
%\end{eqnarray}
%
%This means that when at constant volume, heat is absorved by a system in stable %equilibrium, its temperature is raised.
%
%Similarly, the same condition can be written in terms of the specific heat at %constant volume:
%\begin{eqnarray}
%C_{V} &= & T\frac{\partial S}{\partial T}\arrowvert_{V}=T\frac{\partial %S}{\partial U}\frac{\partial U}{\partial T}\arrowvert_{V}=\frac{\partial %U}{\partial T}\arrowvert_{V},\\
%\frac{\partial^{2}S}{\partial U^{2}}\arrowvert_{V} &= & %-\frac{1}{T^{2}}\frac{\partial T}{\partial %U}\arrowvert_{V}=-\frac{1}{T^{2}}\frac{1}{C_{V}}<0,
%\end{eqnarray}
%\begin{equation}
%C_{V}>0.\label{ts2}
%\end{equation}
The dependence of $C_V$ with the radial coordinate can be calculated from: 
\begin{equation}\label{defcv}
C_{V}=T\left(\frac{dS}{dT}\right)_{V}.
\end{equation}
Using the expressions deduced in the previous sections, we obtain:
\begin{eqnarray}
%\frac{d\rho}{dT}&=&\frac{\rho\left(1+\left[\alpha-(\alpha+1)\left(\frac{\rho}{\rho_{\text{max}}}\right)^{2}\right]\frac{\rho}{\rho_{\text{max}}}\right)}{2\left(\frac{\rho}{\rho_{\text{max}}}\right)\left[\alpha-2(\alpha+1)\left(\frac{\rho}{\rho_{\text{max}}}\right)^{2}\right]}\frac{1}{T},\\
\frac{C_{\text{V}}}{V}&=&\frac{1}{2T}\frac{\rho_{\text{max}}\left(1+\left[\alpha-(\alpha+1)\left(\frac{\rho}{\rho_{\text{max}}}\right)^{2}\right]\frac{\rho}{\rho_{\text{max}}}\right)}{\left[\alpha-2(\alpha+1)\left(\frac{\rho}{\rho_{\text{max}}}\right)^{2}\right]}.
\label{cv2}
\end{eqnarray}
For $r = R$, $\rho$ is zero, and (\ref{cv2}) yields:
\begin{equation}
\frac{C_{\text{Vsup}}}{V_{\text{sup}}}=\frac{\rho_{\text{max}}}{2\alpha T_{\text{sup}}}.
\end{equation}
Introducing this expression as a constant of normalization in (\ref{cv2}), it follows that:
%$$\frac{C_{\text{V}}}{V}\frac{V_{\text{sup}}}{C_{\text{Vsup}}}=\frac{\rho_{\text{max}}\left(1+\left[\alpha-(\alpha+1)\left(\frac{\rho}{\rho_{\text{max}}}\right)^{2}\right]\frac{\rho}{\rho_{\text{max}}}\right)}{\left[\alpha-2(\alpha+1)\left(\frac{\rho}{\rho_{\text{max}}}\right)^{2}\right]}\frac{1}{2T}\frac{2\alpha T_{\text{sup}}}{\rho_{\text{max}}}$$
\begin{equation}
\frac{C_{\text{V}}}{C_{\text{Vsup}}}=\left(\frac{r}{R}\right)^{3}\frac{\alpha\left(1+\left[\alpha-(\alpha+1)\left(\frac{\rho}{\rho_{\text{max}}}\right)^{2}\right]\frac{\rho}{\rho_{\text{max}}}\right)}{\left[\alpha-2(\alpha+1)\left(\frac{\rho}{\rho_{\text{max}}}\right)^{2}\right]}\frac{T_{\text{sup}}}{T}.
\end{equation}
The plot of the specific heat at constant volume is shown in Figure \ref{cv}. We can see that the specific heat is negative for $r/R<0.4$ and positive for $r/R>0.4$. For $r/R=0.4$ it is not defined. The change of sign in $C_{{\rm V}}$ can be understood from Figures \ref{figtemp} and \ref{fig:2}, taking into account that (\ref{defcv}) can also be written as:
\begin{equation}
C_{V}=T\left(\frac{dS}{d\rho}\right)_{V} \left(\frac{d\rho}{d T}\right)_{V}.
\end{equation}

Since the slope of the entropy as a function of the density is always positive, the change of sign is determined by $d\rho / dT$, which can also be written as $d\rho / dT = (d \rho /dr) \; (dr/dT)$. The density as a function of $r$ is a decreasing exponential function so, $d \rho /dr < 0$ for any value of the radius. From Figure \ref{figtemp} we see that for $r/R <0.4$, $dr/dT >0$ and $d\rho /dT$ becomes negative in that region; instead for $r/R>0.4$, $dr/dT <0$ and $d\rho/dT >0$ in the normal matter region.
   
%Figs.\ref{st1} and \ref{st2} display the plot of the entropy as a function of %the temperature. It can be seen in Fig. \ref{st2} that from $T/T_{\text{sup}} = %0$ to $T/T_{\text{sup}} = 3.5$, the slope of the function $S(T)$ is negative %whereas in Fig. \ref{st1} for $T/T_{\text{sup}} = 1$ to $T/T_{\text{sup}} = %3.5$, the slope is positive. These two regions correspond to $r/R \in %\left(0,0.4\right)$ and, $r/R \in \left(0.4,1\right)$. From the thermal %stability conditions written 
%mentioned above, we conclude that only the region of normal matter is in %thermal stable equilibrium.
%\begin{figure}[b]
%\resizebox{\hsize}{!}{\includegraphics{SvsT2.eps}}
%  \end{minipage}~\hfill~\\[0.1pt]
%  \hfill\begin{minipage}[t]{.5\textwidth}
%  \end{minipage}~\hfill%
%  \begin{minipage}[t]{.5\textwidth}
%\caption{Entropy as a function of the temperature in the black hole interior in %the region of normal matter. $S_{T_{\text{sup}}/2}$ stands for the entropy at %$T =T_{\text{sup}}/2$.}
%\label{st1}%
%  \end{minipage}\hfill~%
%\end{figure}
%\begin{figure}[t]
%\resizebox{\hsize}{!}{\includegraphics{SvsT1.eps}}
%  \end{minipage}~\hfill~\\[0.1pt]
%  \hfill\begin{minipage}[t]{.5\textwidth}
%  \end{minipage}~\hfill%
%  \begin{minipage}[t]{.5\textwidth}
%\caption{Entropy as a function of the temperature in the black hole interior in %the region of exotic matter. $S_{T_{\text{sup}}/2}$ stands for the entropy at %$T =T_{\text{sup}}/2$.}
%\label{st2}%
%  \end{minipage}\hfill~%
%\end{figure}
Similarly, we can calculate the specific heat at constant pressure,
defined by:
\begin{equation}\label{defcp}
C_{\text{p}}=T\left(\frac{dS}{dT}\right)_{\text{p}}.
\end{equation}
%\begin{equation}
%C_{P}=V\frac{d\rho}{dT}+(\rho+p)\frac{dV}{dT}-\frac{(\rho+p)V}{T}
%\end{equation}
%Similarly,
%\begin{equation}
%C_{P}/V=\frac{d\rho}{dT}+(\rho+p)\frac{dln(V)}{dT}.
%\end{equation}
%By considering the relation 
%\begin{equation}
%V=4/3\pi r^{3},
%\end{equation}
%we obtain
%\begin{eqnarray}
%\frac{C_{P}/V}{C_{Psup}/V}=&&\left(1-\frac{1}{8r^{3}}(1+[\alpha-(\alpha+1)e^{-16r^{3}/R^{3}}]e^{-8r^{3}/R^{3}})\right)\nonumber\\
%&&\times\frac{\left(1+[\alpha-(\alpha+1)e^{-16r^{3}/R^{3}}]e^{-8r^{3}/R^{3}}\right)}{0.395(\alpha-2(\alpha+1)e^{-16r^{3}/R^{3}})}.
%\end{eqnarray}
%By definition:
%$$C=T\frac{dS}{dT}.$$
%In order to calculate specific heat at constant pressure, we have:
%$$C_{V}=T\left(\frac{dS}{dT}\right)_{P},$$
%considering the equations obtained previously:
%$$S=\frac{(\rho+p)V}{T},$$
%From Eq. (\ref{21}) and considering:
%\begin{equation}
%\left(\frac{dp}{dT}\right)_{\text{p}}=0,
%\end{equation}
%we obtain:
%$$C_{P}=T\left(\frac{\partial \rho}{dT}\frac{V}{T}+\frac{\rho+p}{T}\frac{\partial V}{\partial T}\right)\arrowvert_{P},$$
%\begin{equation}
%C_{\text{p}}/V=\left(\frac{ d\rho}{dT}+(\rho+p)\frac{3}{r}\frac{d r}{d %T}\right)_{\text{p}},
%\end{equation}
The result is:
\begin{equation}
C_{\text{p}}/C_{\text{psup}}=\frac{\alpha}{7}\frac{f_{1}f_{2}}{\left(\frac{T}{T_{\text{sup}}}\right)\left[\alpha-2(\alpha+1)\left(\frac{\rho}{\rho_{\text{max}}}\right)^{2}\right]}_{\text{p}},
\end{equation}
where:
\begin{eqnarray}
f_{1} & = & 8\left(\frac{r}{R}\right)^{3}-1-\frac{\rho}{\rho_{\text{max}}}\left[\alpha-(\alpha+1)\left(\frac{\rho}{\rho_{\text{max}}}\right)^{2}\right],\\
f_{2} & = & 1+\frac{\rho}{\rho_{\text{max}}}\left[\alpha-(\alpha+1)\left(\frac{\rho}{\rho_{\text{max}}}\right)^{2}\right].
\end{eqnarray}
 %\begin{figure}[t]
%\resizebox{\hsize}{!}{\includegraphics{fig8.eps}}
%  \end{minipage}~\hfill~\\[0.1pt]
%  \hfill\begin{minipage}[t]{.5\textwidth}
  
%  \end{minipage}~\hfill%
%  \begin{minipage}[t]{.5\textwidth}
%\caption{Specific heat at constant volume as a function of radial coordinate in the black hole interior. $C_{V{\rm sup}}$ stands for the specific heat at constant volume at $r = R$.}
%\label{cv}%
%  \end{minipage}\hfill~%
%\end{figure}
\begin{figure}[t]
\begin{center}
\includegraphics{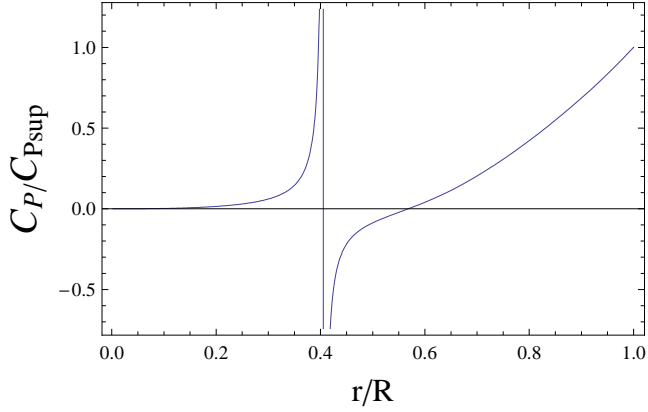}
\end{center}
%  \end{minipage}~\hfill~\\[0.1pt]
%  \hfill\begin{minipage}[t]{.5\textwidth}
  
%  \end{minipage}~\hfill%
%  \begin{minipage}[t]{.5\textwidth}
\caption{Specific heat at constant pressure as a function of the radial coordinate in the black hole interior. $C_{P{\rm sup}}$ stands for the specific heat at constant pressure at $r = R$.}
\label{cp}%
%  \end{minipage}\hfill~%
\end{figure}
%We found that both the specific heat at constant volume and constant pressure are not defined for $r/R= 0.4$, where normal matter changes into exotic matter. The change of sign of $C_{V}$ at $r/R = 0.4$ can be understood in terms of Eq.(\ref{defcv}), the regions where the temperature increases and decreases, and the 
%density profile given in Eq. (\ref{7}). 
The plot of of the specific heat at constant pressure is shown in Figure \ref{cp}. We can see that $C_{p}=0$ at $r/R=0.57$; this point coincides with one of the inflexion points of the temperature. $C_{p}$ is positive for $r/R< 0.4$ and for $r/R>0.57$. In the region $0.4<r/R<0.57$, $C_{p}$ is negative. We write  (\ref{defcp}) as:
\begin{equation}
C_{\text{p}}=T\left(\frac{dS}{dr}\right)_{\text{p}} \left(\frac{dr}{dT}\right)_{\text{p}}.
\end{equation}
From Figure \ref{ENT} we see that $dS/dr >0$ for $r/R<0.57$ and $dS/dr <0$ for $r/R>0.57$ and since the slope of the temperature as a function of the radius is positive for $r/R<0.4$ and negative for $r/R>0.4$, $C_{p}$ presents a change of sign.

The discontinuous behavior of $C_{p}$ as well as of $C_{V}$ at $r/R = 0.4$ is typical of a second-order phase transition, suggesting that the system is thermodynamically unstable. Furthermore, these results show that the specific heats are not defined for the value of $r/R$ where the sound speed equals zero, which reinforces the existence of a region of instability for the normal matter field. The change of sign in both thermodynamical quantities is due to the presence of exotic matter in the core of the object. This is not the case for systems which are only constituted of normal matter.

We arrive at the same conclusion by the examination of the plot of the 
the isothermal compressibility $\kappa_{T}$, defined by:
\begin{equation}
\kappa_{T}=-\frac{1}{V}\left(\frac{\partial V}{\partial p}\right)_{T},
\end{equation}
as a function of $r/R$. The equilibrium condition in this case is
$\kappa_{T} >0$.
%A condition for the system to be in hydrostatic stable equilibrium is that for %a given value of temperature and volume, the Helmholtz free energy should be a %minimum. If the original state of the system is stable, the change in volume %must lead to an increase in the free energy.
%\begin{eqnarray}
%\frac{\partial^{2}F}{\partial V^{2}}\arrowvert_{T} & > &0,\label{h}\\
%\frac{\partial F}{\partial V}& = &-p,\\
%\frac{\partial^{2}F}{\partial V^{2}}\arrowvert_{T} & = & -\frac{\partial %p}{\partial V}\arrowvert_{T}>0,\\
%\frac{\partial p}{\partial V}\arrowvert_{T} & <0 &.
%\end{eqnarray}
%In the latter equation  we see that when the pressure of a stable phase is %increased, the volume must decrease.
%
%We can rewrite the condition (\ref{h}) in terms of $\kappa_{T}$:
%\begin{equation}
%\frac{\partial^{2}F}{\partial V^{2}}\arrowvert_{T}=-\frac{\partial p}{\partial %V}\arrowvert_{T}=\frac{1}{V \kappa_{T}}>0,
%\end{equation}
%which holds if $\kappa_{T} >0$.
\begin{figure}[t]
\begin{center}
{\includegraphics{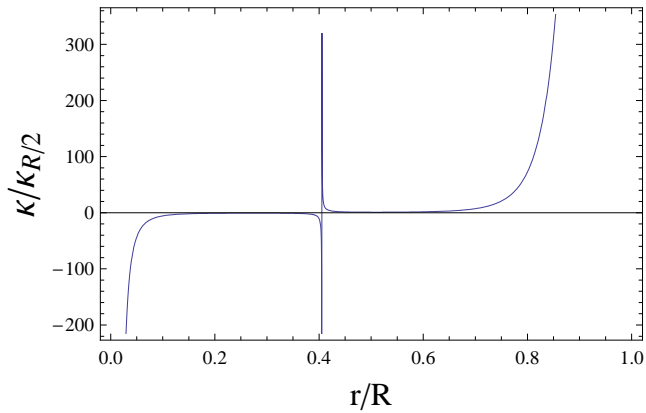}}
\end{center}
%  \end{minipage}~\hfill~\\[0.1pt]
%  \hfill\begin{minipage}[t]{.5\textwidth}
  
%  \end{minipage}~\hfill%
%  \begin{minipage}[t]{.5\textwidth}
\caption{Isothermal compressibility as a function of the radial coordinate. $\kappa_{T R/2}$ stands for the thermal compressibility at $r/R =1/2$.}
\label{ISO}%
%  \end{minipage}\hfill~%
\end{figure}
%%
%%
%
%In the region of exotic matter $\partial^{2}F / \partial V^{2}\arrowvert_{T}  < %0$, while in the region of normal matter fields $\partial^{2}F / \partial %V^{2}\arrowvert_{T}  > 0$. 
%
%It is also shown in Fig. \ref{ISO} the isothermal compressibility $\kappa_{T}$ %as a function of the radial coordinate. As we expected, $\kappa_{T} < 0$ for %$r/R \in \left(0,0.4\right)$ and $\kappa_{T} > 0$ for $r/R \in %\left(0.4,1\right)$. From these results we conclude that the region of exotic %matter is in hydrostatic unstable equilibrium while the region of normal matter %is in hydrostatic stable equilibrium.

\begin{figure}[t]
\begin{center}
{\includegraphics{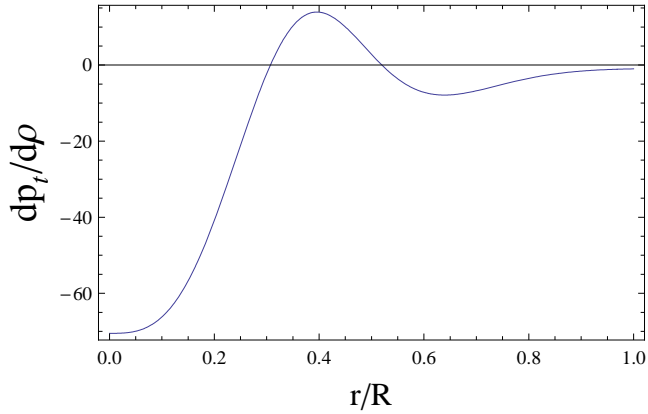}}
\end{center}
%  \end{minipage}~\hfill~\\[0.1pt]
%  \hfill\begin{minipage}[t]{.5\textwidth}
  
%  \end{minipage}~\hfill%
%  \begin{minipage}[t]{.5\textwidth}
\caption{Tangential sound speed as a function of the radial coordinate.}
\label{vtan}%
%  \end{minipage}\hfill~%
\end{figure}

We can summarize the results of this section by asserting that the discontinuities in the second derivatives of the state functions (along with the continuity of the state functions and of their first derivative, as shown in the corresponding plots) indicate that the matter inside the black hole cannot be in thermodynamical equilibrium. This conclusion is reinforced by the plot of the transversal velocity as a function of the $r$, defined by:
$$
v_{\perp}^2 = \frac{dp_{\perp}}{d\rho}.
$$
The function $v_\perp (r)$ can be calculated from 
(\ref{3}) and (\ref{pretan}). This function is plotted in Figure \ref{vtan}.
The plot evidences not only the instability discussed above, but also a new one, inside the normal matter part of the object.

\subsection{Dynamical equilibrium}

To study the dynamical stability of the physical system inside the regular black hole, a detailed analysis 
of the Sturm-Liouville problem associated with the perturbation of the equations of motion for the fluid and the metric is mandatory. However, for the reasons discussed in the Appendix, we shall restrict ourselves here to some arguments suggesting
that the system is dynamically unstable.
 
First, let us recall that in the region of exotic matter there is no propagation of sound waves. The equation for the radial sound speed, in the region of exotic matter, can be put as follows:
\begin{equation}\label{sony}
v_r^{2} = - \frac{\Delta p}{\Delta \rho}.
\end{equation}

%\textbf{DE DONDE VIENE EL SIGNO -?}

From (\ref{sony}) the pressure as a function of the density takes the form:
\begin{equation}\label{sony1}
\Delta p = - v_r^{2} \Delta \rho,
\end{equation}
where $v_r^{2}$ represents the square of the radial sound speed and it is always positive. We can see that if the pressure increases, the density decreases; but if the density decreases, the pressure keeps growing. In this process, both pressure and density are related in such a way by (\ref{sony1}) that, if the system is perturbed, the fluid never stops expanding. We suggest that the huge accumulation of energy in this process of continuous expansion might lead to divergences that indicate the instability of the system.

%...sound waves bounce at $r/R = 0.4$ and accumulate energy in the region between this boundary and the horizon.

%Yet another argument indicating the dynamical instability of the matter configuration follows from the binding energy. 
%Let us recall that in space-times with spherical symmetry,
%the ADM mass $M$ (defined in Eq. (\ref{m})) 
%represents the total energy, that is $M = \mu +$ binding energy, 
%where the proper mass $\mu$ is given by:
%$$
%\mu = 4\pi \int_0^\infty \rho (r)\left(1-\frac{2m(r)}{r}\right)^{1/2}\;r^2\;dr.
%$$
%In the model under scrutiny, $\mu \approx 8.17\times 10^{46} {\rm gr}$, and $M  \approx 4.14\times 10^{31} {\rm gr}$. 

%\textbf{UNIDADES?}

%It follows that the binding energy is positive, hence the system may be unstable, since energy is needed to disassemble it.

\section{Entropy of the gravitational field}
\label{gravent}

Penrose \cite{p2} suggested that entropy might be assigned to the gravitational field itself and he proposed that the Weyl curvature tensor can be used to specify it.

A classical large-scale field such as gravity is expected to have associated an entropy as any other field that can be quantized and obeys the second law of thermodynamics, i.e. tends to a state of equilibrium. Of course, in the absence of a proper quantization of the field we cannot treat it as a gas of gravitons. Instead, we have to rely on approximate classical estimators. Since the equilibrium state of the field seems to be complete gravitational collapse, it appears reasonable to resort to classical estimators based on the Weyl tensor, i.e. the traceless part of the curvature tensor. Recent discussions of these topics are presented by Gr$\varnothing$n \cite{Gronsolo} and by Clifton, Ellis, and Tavakol \cite{Ellis}.

The behavior of the Weyl tensor follows what is expected for a gravitational entropy throughout the history of the universe: it is zero in the (homogeneous) Friedmann-Robertson-Walker model and it is large in Schwarzschild's space-time.

Rudjord, Gr$\varnothing$n and Sigbj$\varnothing$rn \cite{Gron} made a recent attempt to develop a quantitative classical description of the gravitational entropy based on the construction of a scalar derived from the contraction of the Weyl tensor and the Riemann tensor: 
\begin{equation}
P^{2} = \dfrac{C^{\alpha\beta\gamma\delta}C_{\alpha\beta\gamma\delta}}{R^{\alpha\beta\gamma\delta}R_{\alpha\beta\gamma\delta}} = \frac{W}{K}.
\end{equation}

This approach is based on matching their description of the entropy of a black hole in the event horizont with the Hawking-Bekenstein entropy \cite{bek}. In particular, they calculated the entropy of Schwarzschild black holes and the Schwarzschild-de-Sitter space-time \protect\footnote{See reference \cite{us} for other space-times.}.

Rudjord et al. describe the gravitational entropy of a black hole by the surface integral:
\begin{equation}\label{Sint}
S = k_{\rm{s}} \int_\sigma \vec \Psi \cdot \vec{d\sigma},
\end{equation}
where $\sigma$ is the horizon of the black hole and the vector field $\vec{\Psi}$ is:
\begin{equation}
\vec \Psi = P \vec e_{r},
\end{equation}
where $\vec e_{r}$ is a unitary vector in the radial direction. They required (\ref{Sint}) to coincide at the horizon with the Hawking-Bekenstein entropy,
%\begin{equation}
%S=S_{\rm{HB}}.
%\end{equation} 
thus allowing for the calculation of the constant $k_{\rm{s}}$\protect\footnote{To simplify the notation, we shall set the constant $k_{\rm{s}}$ equal to 1 since it plays in what follows the role of a scale factor.}.

Finally, the entropy density can be determined by means of Gauss's divergence theorem, rewriting (\ref{Sint}) as a volume integral:
\begin{equation}\label{den}
\mathfrak{s} = k_{\rm{s}}\mid{ \nabla \cdot \vec \Psi}\mid,
\end{equation}
where the absolute value brackets were added to avoid negative values of entropy.

Let us recall that the Weyl tensor 
%is a 4-rank tensor that contains the independent components of the Riemann %tensor not captured by the Ricci tensor. It can be considered as 
is the traceless part of the Riemann tensor, and is given by:
\begin{eqnarray}\label{weyl-for}
C_{\alpha\beta\gamma\delta} &=& R_{\alpha\beta\gamma\delta}+\frac{2}{n-2}(g_{\alpha[\gamma}R_{\delta]\beta}-g_{\beta[\gamma}R_{\delta]\alpha})+\\ \nonumber
& + & \frac{2}{(n-1)(n-2)} R~g_{\alpha[\gamma}g_{\delta]\beta},
\end{eqnarray}
where $R_{\alpha\beta\gamma\delta}$ is the Riemann tensor, $R_{\alpha\beta}$ is the Ricci tensor, $R$ is the Ricci scalar, $[\ ]$ refers to the antisymmetric part, and $n$ is the number of dimensions of space-time.

The absence of structure in space-time corresponds to a null Weyl conformal curvature ($W=C^{abcd}C_{abcd}=0$). Hence, the Weyl tensor contains the information of the gravitational field in absence of matter and other non-gravitational fields.  

The Weyl scalar for the space-time metric given in (\ref{met}) is:
\begin{eqnarray}
W &=& \frac{4}{3}\pi^2\rho^2_{\texttt{max}}(p_{\small{8}}\:x^8+p_{\small{6}}\:x^6+p_{\small{5}}\:x^5+p_{\small{4}}\:x^4+p_{\small{3}}\:x^3+p_{\small{2}}\:x^2+ \nonumber \\
& + & p_{\small{1}}\:x+p_{\small{0}})^{2} / \left[\left(3r-\pi R^3\rho_{\rm{max}}+\pi R^3 \rho_{\rm{max}} \:x\right)r^6R^6\right],
\end{eqnarray}
where $x= e^{-8\frac{r^{3}}{R^{3}}}$ and the coefficients $p_{\rm{i}}$ are:
\begin{eqnarray}
p_{\small{0}} & = & 3R^6r-R^9\pi\rho_{\rm{max}}, \nonumber \\
p_{\small{1}} & = & 2R^9\pi\rho_{\rm{max}}-3R^6r+6r^3R^6\pi\rho_{\rm{max}}-24r^4R^3, \nonumber \\
p_{\small{2}} & = & -6r^3R^6\pi\rho_{\rm{max}}-R^9\pi\rho_{\rm{max}}+576r^7\alpha- \nonumber \\
& - & 192r^6\alpha\pi R^3\rho_{\rm{max}}-2R^6r^3\alpha \pi \rho_{\rm{max}}, \nonumber \\
p_{\small{3}} & = & 2 R^6 r^3\alpha\pi\rho_{\rm{max}}+144r^6\alpha\pi R^3\rho_{\rm{max}}, \nonumber \\
p_{\small{4}} & = & (-1152-1152\alpha)r^7-48\pi R^3\rho_{\rm{max}}(-8\alpha-8+\alpha^2)r^6+ \nonumber \\
& + & 2\pi\rho_{\rm{max}} R^6(\alpha+1)r^3, \nonumber \\
p_{\small{5}} & = & -336\pi\rho_{\rm{max}} R^3(\alpha+1)r^6-2\pi\rho_{\rm{max}} R^6(\alpha+1)r^3, \nonumber \\
p_{\small{6}} & = & 96\pi\rho_{\rm{max}}\alpha R^3(\alpha+1)r^6, \nonumber \\
p_{\small{8}} & = & -48 \pi R^3\rho_{\rm{max}}(\alpha+1)^2r^6.
\end{eqnarray}

If we let $r \rightarrow 0$, the Weyl scalar goes to zero. Since the regular black hole space-time has a de Sitter geometry in the origin, this is the expected limit.

\begin{figure}[t]
\begin{center}
{\includegraphics{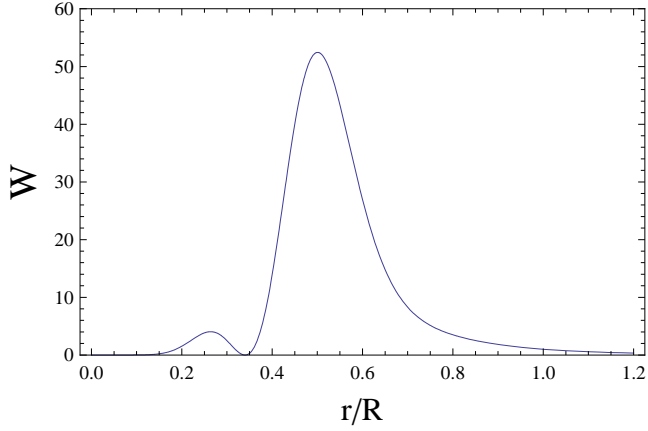}}
\end{center}
\caption{Weyl scalar as a function of the radial coordinate. \label{Weyl}}
\end{figure}

%\begin{figure}[pb]
%\centerline{\psfig{file=fig11.eps,width=4.7cm}}
%\vspace*{8pt}
%\caption{Kretschmann scalar as a function of radial coordinate. \label{Kre}}
%\end{figure}

\begin{figure}[t]
\begin{center}
{\includegraphics{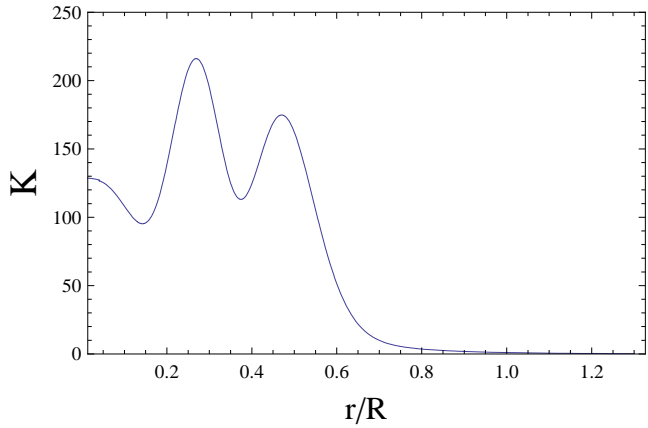}}
\end{center}
%  \end{minipage}~\hfill~\\[0.1pt]
%  \hfill\begin{minipage}[t]{.5\textwidth}
  
%  \end{minipage}~\hfill%
%  \begin{minipage}[t]{.5\textwidth}
\caption{Kretschmann scalar as a function of the radial coordinate.}
\label{Kre}%
%  \end{minipage}\hfill~%
\end{figure}

%\begin{figure}[b]
%  \centering
%  \hfill\begin{minipage}[b]{.45\textwidth}
%  \centering
%  \includegraphics[width=\textwidth]{Wyes.eps}
%  \end{minipage}~\hfill%
%  \begin{minipage}[b]{.45\textwidth}
%    \centering
%    \includegraphics[width=\textwidth]{Kyes.eps}
%  \end{minipage}~\hfill~\\
%  \hfill\begin{minipage}[t]{.45\textwidth}
%  \caption{Weyl scalar as a function of radial coordinate.}
%  \label{Weyl}%
%  \end{minipage}~\hfill%
%  \begin{minipage}[t]{.45\textwidth}
%    \caption{Kretschmann scalar as a function of radial coordinate.}
%    \label{Kre}%
%  \end{minipage}\hfill~%
%\end{figure}

The plot of the Weyl scalar is shown in Figure \ref{Weyl}. We see that for large values of $r$ the Weyl scalar tends asymptotically to zero. In the matter region ($r/R < 1$) it has one absolute maximum at $r/R = 0.5$ and one relative maximum at $r/R = 0.26$. Between these points, $W=0$ at $r/R = 0.34$. If we analyze the equation of state as a function of the density (Equation (\ref{3})), we find one inflexion point at $r/R = 0.5$. The absolute maximum of the Weyl scalar seems to be related to the transition point between the two regions of matter.

The value of $r/R$ for which the Weyl scalar has one relative maximum is close to the point where the pressure is zero. This suggests a relation between the region where matter has a negative pressure and the behavior of the Weyl scalar. Notice that the inflexion point of the entropy density of the matter coincides with the value of $r/R$ for which the Weyl scalar is zero, that is at $r/R= 0.34$. At this point the gravitational field changes from attractive to repulsive.

We also calculate the Kretschmann scalar. The plot is shown in Figure \ref{Kre}. Again, we see that outside the matter region the Kretschmann scalar tends asymptotically to zero. If we let $r\rightarrow0$, it goes to:
\begin{equation}
K = \frac{512\pi^2{\rho_{\rm{max}}}^2}{3},
\end{equation}
as expected for the de Sitter space-time. The Kretschmann scalar is always positive and it has one absolute maximum at $r/R = 0.27$ and one relative maximum at $r/R = 0.47$.

As a cross-check, we see from the plots that the calculated Weyl and Kretschmann scalars satisfy the relation:
\begin{equation}
R^{\alpha \beta \gamma \delta} R_{\alpha \beta \gamma \delta} \geq C^{\alpha \beta \gamma \delta} C_{\alpha \beta \gamma \delta},
\end{equation}
that holds in every static spherically symmetric space-time \cite{Gron}.
%\begin{figure}[t]
% \centering
%  \hfill\begin{minipage}[b]{.45\textwidth}
%\includegraphics[scale=0.3]{entroyes.eps}
%\includegraphics{entropia-final.eps}
%\caption{Gravitational entropy density as a function of radial coordinate.}
%\label{entropy-grav}%
%\end{figure}
%\begin{figure}[pt]
%\centerline{\psfig{file=fig12.eps,width=4.7cm}}
%\vspace*{8pt}
%\caption{ \label{entropy-grav}}
%\end{figure}
\begin{figure}

\resizebox{\hsize}{!}{\includegraphics{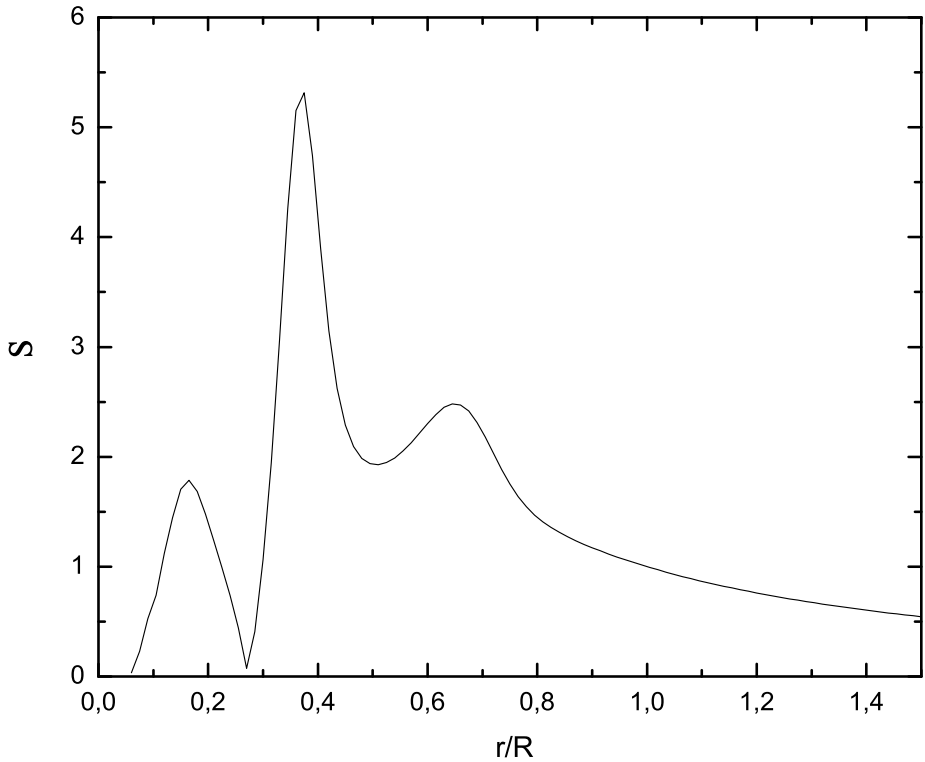}}
%{\includegraphics{fig16.eps}}
%  \end{minipage}~\hfill~\\[0.1pt]
%  \hfill\begin{minipage}[t]{.5\textwidth}
  
%  \end{minipage}~\hfill%
%  \begin{minipage}[t]{.5\textwidth}
\caption{Gravitational entropy density as a function of the radial coordinate.}
\label{entropy-grav}%
%  \end{minipage}\hfill~%
\end{figure}
%\begin{figure}[pt]
%\centerline{\psfig{entropia-final.eps,width=4.7cm}}
%\vspace*{8pt}
%\caption{Gravitational entropy density as a function of radial coordinate. \label{entropy-grav}}
%\end{figure}
The gravitational entropy density of the MK model can be computed following following Rudjord et. al.'s proposal.  The plot of the entropy density is shown in Figure \ref{entropy-grav}. It is seen that $\mathfrak{s} = 0$ at $r/R = 0.27$, which is close to the transition point from positive to negative pressures. The absolute maximum is at $r/R = 0.37$ and the two relative maxima are at $r/R = 0.16$ and at $r/R = 0.65$. For large values of $r/R$ the gravitational entropy density tends asymptotically to zero and in the core of the object it is also zero. This last result is in accordance with a correct classical description of the entropy of the gravitational field.

\section{Conclusions}
\label{disc}

We have shown that the thermodynamical quantities describing 
the matter that is the source of the regular black hole model proposed by Mbonye and Kazanas indicate that the region of exotic matter is unstable. Some evidence has been presented that points to the dynamical instability of the model as well. 
%In a recent work, E. Trojan and G. Vlasov \cite{trojan} study the %thermodynamics of exotic matter by means of a statistical analysis of an ideal %Fermi gas at zero temperature. They also find the same thermodynamic properties %for the exotic matter field described in this work.
The evidence for this second instability is supported by the plots of $v_r$ and $v_\perp$. In particular, the latter indicates the existence of a second type of dynamical instability, in the normal matter
part of the object. These findings should be confirmed by a perturbation
analysis (which will be presented elsewhere), based on the equations given in the Appendix.

We have also calculated the Weyl and Krestchmann scalars for the regular MK black hole, and a possible classical estimator of the gravitational entropy proposed by Rudjord, Gr$\varnothing$n and Sigbj$\varnothing$rn based on the Weyl curvature conjecture. It was shown that close to the core of the object, the entropy density tends to zero for large values of the radius. Hence, 
this classical estimator gives a good description of the entropy of the gravitational field in de Sitter and Schwarzschild limits.

%\begin{acknowledgements}
%BH astrophysics with G.E. Romero is supported by grant PIP 2010/0078 (CONICET). Additional funds comes from Spanish grant AYA 2010-21782-C03-01. SEPB acknowledges support from UERJ, FAPERJ, anc CNPQ. We are grateful to Camila Correa for early contributions to this research.
%\end{acknowledgements}

\begin{acknowledgements}
BH astrophysics with G.E. Romero is supported by the Argentine agencies ANPCyT (BID 1728/OC -
AR PICT-2012-00878) and CONICET (PIP 0078/2010), as well as the Spanish grant AYA
2010-21782-C03-01. SEPB would like to acknowledge support from UERJ, FAPERJ, and ICRANet-Pescara. We are grateful to Camila A. Correa for early contributions to this research.

\end{acknowledgements}

\appendix
\section*{Appendix: dynamical stability of the interior region}

In order to study the stability of a system with tangential pressures
against radial perturbations, both Einstein field equations and the equation of state must be perturbed following the standard procedure (see for instance \cite{hille}). The goal is to obtain a differential equation for the radial dependence of the quantity $\xi(r,t)$,
% by a quantity $\xi(r,t)= \sigma(r) \exp{i \omega t}$ 
which represents a small radial displacement of the fluid respect to its position of equilibrium at time $t$, that is, $r(r,t)=r_{0}+\xi(r,t).$

In what follows, the unperturbed metric coefficients $A(r)$, $B(r)$, and the unpertubed thermodynamics variables 
%$p_{\text{r}}(r,t)=p_{\text{r}}$, $p_{\perp}(r,t)=p_{\perp}$, and %$\rho(r,t)=\rho$ 
will be denoted by a zero subindex.
The corresponding perturbations are denoted by $\delta A(r,t)$, $\delta B(r,t)$, $\delta p_{\text{r}}(r,t)$, $\delta p_{\perp}(r,t)$ and $\delta \rho(r,t)$.
A long but straightforward calculation
(which parallels that presented in \cite{hille})
shows that
the equations for the perturbations are given by:
%the perturbations of the field equations results in:
%
\begin{eqnarray}
8\pi r^{2}\delta\rho & = & \frac{\partial}{\partial r}\left(\frac{r\delta A}{A_{0}^{2}}\right),\\\label{6N}
8\pi\dot{\xi}(\rho_{0}+p_{\text{r}}^{0}) & = & -\frac{1}{r}\frac{\partial \delta A}{\partial t}\frac{1}{A_{0}^{2}},\\\label{7N}
8\pi r^{2}\delta p_{\text{r}} & = & \frac{r}{A_{0}}\left[\frac{\partial}{\partial r}\left(\frac{\delta B}{B_{0}}\right)-\frac{dB_{0}}{dr}\frac{\delta A}{A_{0}}\right]-\frac{\delta A}{A_{0}^{2}},\\\nonumber
\delta p_{\perp} & = & \frac{r}{2}\frac{\partial \delta p_{\text{r}}}{\partial r}+\delta p_{\text{r}}+\frac{r}{4B_{0}}\frac{dB_{0}}{dr}(\delta p_{\text{r}}+\delta \rho)\\
& & +\left[\frac{r}{4}\frac{\partial}{\partial r}\left(\frac{\delta B}{B_{0}}\right)+\frac{r}{2}\frac{A_{0}}{B_{0}}\frac{\partial^{2}\xi}{\partial t^{2}}\right](p_{\text{r}}^{0}+\rho_{0}).
\end{eqnarray}
Notice that there are six unknowns in this system, a consequence 
of the choices made by MK (one EOS and the explicit form of 
$\rho (r)$ to determine the system).

Equation (\ref{6N}) can be integrated respect to time \footnote{We choose the constant of integration so that $\delta A(r,t)=0$ if $\delta r=0$.}, yielding:
\begin{equation}
\frac{\delta A}{A_{0}}=-8\pi r(\rho_{0}+p_{\text{r}}^{0})A_{0}\xi.
\end{equation}
By isolating $\delta B$ from (\ref{7N}) we get:
\begin{equation}
\frac{\partial}{\partial r}\left(\frac{\delta B}{B_{0}}\right)=8\pi rA_{0}\delta p_{\text{r}}+\frac{\delta A}{A_{0}}\left(\frac{1}{B_{0}}\frac{dB_{0}}{dr}+\frac{1}{r}\right).
\end{equation}
The final set of equations takes the form:
\begin{eqnarray}\label{10}
8\pi r^{2}\delta\rho & = & \frac{\partial}{\partial r}\left(\frac{r\delta A}{A_{0}^{2}}\right),\\\label{11}
\delta p_{\text{r}}+(p_{\text{r}}^{0})'\xi & = & \left(\frac{dp_{\text{r}}}{d\rho}\right)_{0}(\delta \rho+\rho_{0}'\xi),\\\label{12}
\delta p_{\perp} & = & \frac{r}{2}\frac{\partial \delta p_{\text{r}}}{\partial r}+\delta p_{\text{r}}+\frac{r}{4B_{0}}\frac{dB_{0}}{dr}(\delta p_{\text{r}}+\delta \rho)\\
\nonumber
& & +\left(\frac{r}{4}\frac{\partial}{\partial r}\left(\frac{\delta B}{B_{0}}\right)+\frac{r}{2}\frac{A_{0}}{B_{0}}\frac{\partial^{2}\xi}{\partial t^{2}}\right)(p_{\text{r}}^{0}+\rho_{0}),\\\label{13}
\frac{\delta A}{A_{0}} & = & -8\pi r(\rho_{0}+p_{0})A_{0}\xi,\\\label{14}
\frac{\partial}{\partial r}\left(\frac{\delta B}{B_{0}}\right) & = & 8\pi rA_{0}\delta p_{\textsf{r}}+\frac{\delta A}{A_{0}}\left(\frac{1}{B_{0}}\frac{dB_{0}}{dr}+\frac{1}{r}\right).
\end{eqnarray}
Standard manipulations using the expression 
$\xi(r,t)= \sigma(r) \exp{i \omega t}$ in these equations
lead to the following second order ordinary inhomogeneous differential equation for  the function $\sigma(r)$:
\begin{equation}\label{SL}
A_{\star}(r)\sigma''(r)+C_{\star}(r)\sigma'(r)+\left[B_{\star}(r)+\omega D_{\star}(r)\right]\sigma(r)=
\Pi(r),
\end{equation}
where we have also used $\delta p_{\perp}(r,t) = \Pi (r)\exp{i \omega t}$.

%\textbf{REVISAR, ESTO ERA ASI, VERDAD?}

This equation defines an \textit{inhomogeneous} Sturm-Liouville (SL) problem, 
differently from the common case which yields a homogeneous one, the inhomogeneity being a direct consequence of the way the MK model solution was obtained, as mentioned before. 

\begin{figure}
\begin{center}
{\includegraphics{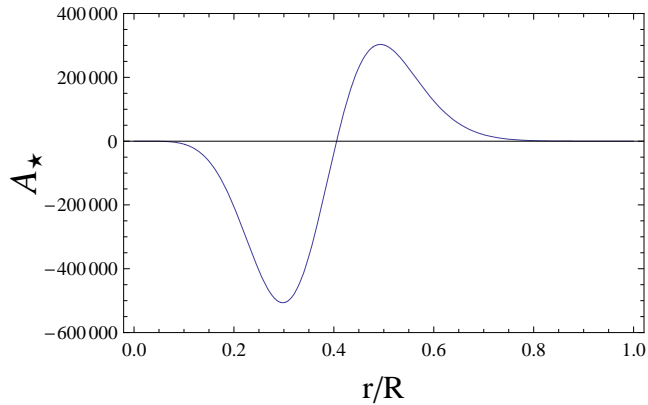}}
\end{center}
%  \end{minipage}~\hfill~\\[0.1pt]
%  \hfill\begin{minipage}[t]{.5\textwidth}
  
%  \end{minipage}~\hfill%
%  \begin{minipage}[t]{.5\textwidth}
\caption{Coefficient $A_{\star}$ as a function of the radial coordinate.}
\label{coef-A}%
%  \end{minipage}\hfill~%
\end{figure}

Equation (\ref{SL}) is to be solve for $r \in \left[0,1\right]$, using the boundary conditions
%\begin{equation}
%\sigma (0) = 0,
%\end{equation}
%\begin{enumerate}
%$\item 
%$\begin{equation}\label{bc1}
%$\xi(0,t) =0 \ \ \ \ \ \Rightarrow \ \ \ \ \ \sigma(r) =0.
%$\end{equation}      
%$\item 
%$\begin{equation}
%$\Delta p_{{\rm r}}(1,t)= \left(\frac{\partial p_{{\rm r}(1,t)}}{\partial %$\rho}\right) \left[\delta \rho (1,t)+\rho'_{0}(1,t) \xi(1,t)\right] =0.
%\end{equation}
%
%\begin{equation}
%\left[\delta \rho (1,t)+\rho'_{0}(1,t) \xi(1,t)\right] =  0.
%\end{equation}
%
%After some calculations, we get:
%\begin{equation}
%\sigma(1) = \epsilon \sigma'(1),
%\end{equation}
%where $\epsilon = -2.37 \:10^{-92}$. Since $\epsilon \approx 0$, our second boundary condition is:
%\begin{equation}\label{bc2}
%\sigma(1) = 0.
%\end{equation}
%\end{enumerate}
$\sigma(r) =0$, $\sigma(1) = 0$, and $\left[\delta \rho (1,t)+\rho'_{0}(1,t) \xi(1,t)\right] =  0$ \cite{hille}.
%Using formulas (\ref{bc1}) and (\ref{bc2}) we can show that we have separated %boundary conditions.
The coefficients $A_{\star}(r)$, $B_{\star}(r)$, $C_{\star}(r)$, and $D_{\star}(r)$ are given by:
\begin{equation}
A_{\star}(r) = \frac{r}{2} \left(p_{\text{r}}^{0}+\rho_{0}\right)\left(\frac{dp}{d\rho}\right)_{0} ,
\end{equation}
\begin{equation}
B_{\star}(r) = B_{1}(r)+B_{2}(r)+B_{3}(r)+B_{4}(r),
\end{equation}
where:
\begin{eqnarray}
B_{1}(r) & = & \frac{1}{8\pi r^2}q'(r)\alpha, \\
q'(r) & = & \frac{\partial}{\partial r}\left[-8\pi r^{2} \left(\rho_{0}+p_{\text{r}}^{0}\right)\right],\nonumber\\
\alpha & = & \frac{r}{2} \frac{\partial}{\partial r}\left(\frac{dp}{d\rho}\right)_{0}+\left(\frac{dp}{d\rho}\right)_{0} \Sigma(r) +\frac{r}{4B_{0}} \frac{dB_{0}}{dr},\nonumber \\
\Sigma(r) & = & 1+ \frac{r}{4B_{0}} \frac{dB_{0}}{dr}+4 \pi r^2 A_{0} \left(p_{\text{r}}^{0}+\rho_{0}\right),\nonumber 
\end{eqnarray}
\begin{eqnarray}
B_{2}(r) & = & \frac{r}{2} \tau(r), \\
\tau(r) & = & \frac{\partial}{\partial r}\left(\frac{dp}{d\rho}\right)_{0} \rho'_{0}-\frac{\partial}{\partial r}\left(p_{\text{r}}^{0}\right)'+ \left(\frac{dp}{d\rho}\right)_{0} \frac{\partial \rho'_{0}}{\partial r},\nonumber 
\end{eqnarray}
\begin{equation}
B_{3}(r)  =  \frac{r}{2} \frac{\partial}{\partial r}\left(\frac{1}{8 \pi r^2} q'(r)\right)\left(\frac{dp}{d\rho}\right)_{0},
\end{equation}
\begin{equation}
B_{4}(r)  =  - 8 \pi r \left(\rho_{0}+p_{\text{r}}^{0}\right) A_{0} \left[r \frac{d}{dr}\left(\ln B_{0}\right) + 1\right],
\end{equation}
\begin{equation}
C_{\star}(r) = C_{1}(r)+C_{2}(r)+C_{3}(r),
\end{equation}
\begin{eqnarray}
C_{1}(r)& = &\frac{1}{8 \pi r^{2}}q \alpha,\\
q & = & -8\pi r^{2} \left(\rho_{0}+p_{\text{r}}^{0}\right),\nonumber 
\end{eqnarray}
\begin{equation}
C_{2}(r) = \frac{r}{2} \left[\left(\frac{dp}{d \rho}\right)_{0} \rho'_{0}-p'_{0}\right],
\end{equation}
\begin{equation}
C_{3}(r) = \frac{1}{8 \pi r}\left(-\frac{1}{r}q+ q'\right),
\end{equation}
\begin{equation}
D_{\star}(r) = - \frac{r}{2} \frac{A_{0}}{B_{0}} \left(p_{\text{r}}^{0}+\rho_{0}\right).
\end{equation}
For a given $\Pi (r)$, (\ref{SL}) might be solved using the Green's function method (see for instance \cite{morse}). However, as we shall show next, the
SL problem defined by (\ref{SL}) has a singular point. 
Consider the equation \cite{zettl}:
\begin{equation}
-\left(py'\right)'+qy= \lambda wy, 
\end{equation}
which corresponds to the homogeneus SL problem defined on 
on $J = (a,b)$, with $-\infty \leq a \leq b \leq \infty$.
The eigenvalue $\lambda $ is such that $\lambda \in \mathbb{C}$, and $p,q,w$ and $y$ are functions of $x$. 
It is also assumed that:
\begin{equation}
\frac{1}{p}, q, w \in L_{{\rm loc}}\left(J,\mathbb{C}\right),
\end{equation}
where $L_{{\rm loc}}\left(J,\mathbb{C}\right)$ denotes the linear manifold of functions $y$ satisfying $y \in L\left(\left[\alpha,\beta\right],\mathbb{C}\right)$ for all compact intervals $\left[\alpha,\beta\right] \subseteq J$.
$L\left(J,\mathbb{C}\right)$ denotes the linear manifold of complex valued Lebesgue measurable functions $y$ defined on $J$ for which:
$\int_{a}^{b} \left|y(t)\right| dt \equiv \int_{J} \left|y(t)\right| dt \equiv \int_{J} \left|y(t)\right| < \infty$.\\
Here $J$ is any interval of the real line, open, closed, half open, bounded or unbounded.
Following \cite{zettl}, we have the next definitions:
\begin{itemize}
\item The (finite or infinite) endpoint $a$ is \textsl{regular} if
\begin{equation}
\frac{1}{p}, q, w \in L((a,d),\mathbb{C}),
\end{equation}
holds for some (hence any) $d \in J$.
\item An endpoint is called \textsl{singular} if it is not regular,
\end{itemize}
with similar definitions at $r=b$.
To analyse whether $a=0$ is a regular or singular endpoint for our problem, defined by (\ref{SL}),
we show in Figure \ref{coef-A} the plot of the coefficient $A_{\star}$ as a function of radial coordinate.
We can see that $A_{\star}(0)= A_{\star}(0.4)=0$, $A_{\star}(r)<0$ for $r \in (0,0.4)$, and $A_{\star}(r)>0$ for $r \in (0.4,1)$. Hence, $\frac{1}{A}_{\star} \notin L((0,d),\mathbb{C})$ for any $d \in (0,1)$. Therefore, the endpoint $a =0$ is singular. 
If an endpoint is singular it could be a \textit{limit point} or a \textit{limit circle}. 
According to \cite{zettl}, ``there is hardly any literature on the LP/LC (limit point-limit circle) dichotomy when all three
coefficients are present in the SL equation''. In particular, 
``there seems to be no literature on LP/LC criteria when $p$ changes sign''.  
Due to these complications, we shall attack this problem in a future 
publication.
%A. Zettl in his book \textsl{Sturm-Liouville Theory}, page 152\cite{zettl}:
%\begin{quote}
%There is hardly any literature on the LP/LC (limit point-limit circle) %dichotomy when all three
%coefficients are present. Most of the literature just considers the half line %case $J=(a,\infty)$ with a a finite regular endpoint, and with $w = 1$. A few %authors have studied the case when both $p$ and $q$ are present and $p > 0$ %case but most just set $p = 1$. \textbf{There seems to be no literature on %LP/LC criteria when $p$ changes sign}.  
%\end{quote}

%The classification of the endpoints $a=0$ and $b=1$ in order to characterize a%nd solve the Sturm-Liouville Eq. (\ref{SL}) remains an open problem that we %will study in future work. It is possible, however, by means of a qualitative %physical argument to show that this particular case of a regular black hole %model is unstable.

%\newpage
%\section*{References}

\end{document}